\documentclass[prd,preprint,eqsecnum,amsmath,amssymb,nofootinbib]{revtex4}
\usepackage{graphicx, tikz}% Include figure files
\usepackage{caption, subcaption}
\usepackage{amssymb,amsmath, mathrsfs}
\usepackage{amssymb, graphics, setspace}
\usepackage{hyperref}
\usepackage[all]{xy}
\newcommand{\be}{\begin{equation}}
\newcommand{\ee}{\end{equation}}
\newcommand{\bea}{\begin{eqnarray}}
\newcommand{\eea}{\end{eqnarray}}
\newcommand{\bes}{\begin{subequations}}
\newcommand{\ees}{\end{subequations}}

\begin{document}

\title{The Hawking effect in the particles-partners correlations\footnote{This paper is dedicated to te memory of Francesco Saverio Persico which one of the authors (R.B.) had the priviledge to know during his stay at the University of Palermo.}}
%region of BHs \\ in 1+1 dimensional acoustic analogs}
\author{Roberto~Balbinot}
\email{roberto.balbinot@unibo.it}
\affiliation{Dipartimento di Fisica dell'Universit\`a di Bologna and INFN sezione di Bologna, Via Irnerio 46, 40126 Bologna, Italy
%\\ Centro Fermi - Museo Storico della Fisica e Centro Studi e Ricerche Enrico Fermi, Piazza del Viminale 1, 00184 Roma, Italy
%
}
\author{Alessandro~Fabbri}
\email{afabbri@ific.uv.es}
\affiliation{Departamento de F\'isica Te\'orica and IFIC, Universidad de Valencia-CSIC, C. Dr. Moliner 50, 46100 Burjassot, Spain
%\\  Centro Fermi - Museo Storico della Fisica e Centro Studi e Ricerche Enrico Fermi, Piazza del Viminale 1, 00184 Roma, Italy  \\
%Laboratoire de Physique Th\'eorique, CNRS UMR 8627, B\^at. 210, Universit\'e Paris-Sud 11, 91405 Orsay Cedex, France
}

\bigskip\bigskip

\begin{abstract}

We analyze the correlations functions across the horizon in Hawking black hole radiation to reveal the correlations between Hawking particles and their partners. The effects of the underlying space-time on this are shown in various examples ranging from acoustic black holes to regular black holes.  

\end{abstract}
\maketitle

\section{Introduction}

The 1974 prediction by Hawking \cite{hawking} of a quantum thermal emission by black holes (BHs) is a milestone of modern theoretical physics. Being the associated temperature extremely tiny (of order $10^{-8}\ K$ for a solar mass BH) no experimental evidence of this remarkable result has been given so far. Nowadays the only indication that this phenomenon can indeed exist in Nature comes surprisingly from condensed matter physics. An analog of a BH created by a Bose-Einstein condensate (BEC) undergoing a transition from a subsonic flow to a supersonic one has shown \cite{jeff1, jeff2, jeff3}, in the density-density correlation function across the horizon, a characteristic peak \cite{paper1, cfrbf} which is consistent with a pair-creation mechanism of a Hawking particle outside the horizon and its entangled partner inside. This has stimulated a large interest in studying quantum correlations across the horizon in relation to the Hawking effect.

In this paper we shall consider various BH metrics and analyze if and where this kind of signal does indeed appear. We shall simplify the mathematical treatment by considering two-dimensional (2D) spacetimes (some of them can be considered as the time-radial section of spherically symmetric 4D spacetimes) and a massless scalar quantum field propagating on them. We shall focus on a particular component of the quantum energy momentum tensor of the scalar field which is relevant to reveal the presence of the Hawking effect. 

\section{The setting}
\label{s2}

The 2D BH metrics we shall consider are stationary and can be cast in the Eddington-Finkelstein (EF) form
\be \label{dueuno} ds^2=-f(r)dv^2+2dvdr \ , \ee
where $v$ is a null advanced coordinate. The horizon $r_H$ corresponds to $f(r_H)=0$. A retarded coordinate $u$ can be introduced by 
\be \label{duedue} u=v-2r_* \ , \ee
where the Regge-Wheeler coordinate $r_*$ is
\be \label{duetre}
r_*=\int \frac{dr}{f(r)}\ . \ee
One can write the metric in the double-null form 
\be \label{duequattro}
ds^2=-f(r)dudv \ , \ee
where $r$ is a function of $u$ and $v$ defined implicitly by
\be \label{duecinque}
r_*=\frac{v-u}{2}\ . 
\ee 

We consider a massless scalar field $\hat \phi(x)$ propagating in the spacetime of eq. (\ref{dueuno}) (or (\ref{duequattro})), satisfying 
\be \hat\Box\hat\phi (x)=0 \ , \label{dueotto} \ee
where $\hat\Box\equiv \nabla_\mu\nabla^\mu$ is the covariant d'Alembertian and $x$ is a generic space-time point. 
The energy momentum operator associated to $\hat\phi$ reads
\be \label{duenove} 
\hat T_{ab}(\hat \phi(x)) = \partial_a \hat\phi(x)\partial_b\hat\phi(x) -\frac{g_{ab}}{2}\partial^c\hat\phi(x)\partial_c\hat\phi(x)\ . \ee
The (across the horizon)  correlator of this operator we will study is \cite{Parentani:2010bn}
\be \label{duedieci} G(x,x')=\frac{\langle U|\hat T_{uu}(\hat\phi (r,v))\hat T_{u'u'}(\hat \phi(r',v')|U\rangle }{f^2(r)f^2(r')}\ , \ee
where $r>r_H$ and $r'<r_H$.

In the double null coordinate system of eq. (\ref{duequattro}) 
\be \label{dueundici}
\hat T_{uu} (\hat\phi) =\partial_u\hat\phi\partial_u\hat\phi \ .
\ee
The quantum state $|U\rangle$ in which the expectation value in eq. (\ref{duedieci}) is taken is the Unruh state \cite{Unruh:1976db}. This state is defined by expanding the field $\hat\phi$ in a base
\be \label{duedodici} \{ \frac{e^{-i\omega_kU}}{\sqrt{2\pi\omega_k}}\ ; \frac{e^{-i\omega v}}{\sqrt{2\pi\omega}} \} 
\ , \ee
where $U$ is Kruskal coordinate defined 
\be \label{duetredici} U=\mp \frac{1}{\kappa}e^{-\kappa u} \ , \ee
which, unlike $u$, is regular on the future horizon. In eq. (\ref{duetredici}) $\kappa$ is the surface gravity of the horizon 
\be \label{duequattordici} \kappa=\frac{1}{2} \frac{df}{dr}|_{r_H} \ee
and $-$ holds outside the horizon, while $+$ inside. 
The state $|U\rangle$ describes the state of the quantum field at late retarded time $u$ after the BH is formed and is the relevant one to discuss the Hawking BH evaporation in this limit. 

The correlator eq. (\ref{duedieci}) is quite general. For a gravitational black hole it is related to the energy density correlator measured by geodesic observers. Its square root appears in acoustic BH giving the density-density correlator \cite{paper1, cfrbf} in a BEC under the hydrodynamical approximation.\footnote{In this case the equal-time condition is taken at Painlev\'e time (the laboratory time). The qualitative features we shall discuss remain unchanged.}
The two-point function for the $u$ sector of the state $|U\rangle$ is 
\be \label{duequindici} 
\langle U| \hat \phi(x)\hat\phi (x') |U\rangle =-\frac{\hbar}{4\pi} \ln(U-U') \ee
from which one easily gets 
\be G(r,r')=\frac{1}{f^2(r)f^2(r')}\left( \frac{\hbar\kappa^2}{16\pi\cosh^2[\frac{\kappa}{2}(u-u')]} \right)^2|_{v=v'}\ , \label{duesedici} \ee
where \be \label{duediciassette} (u-u')|_{v=v'}=-2(r_*-r_*')\ . \ee
This function is expected to display the correlations of the particle-partner pairs. The two, when on-shell, propagate along $u=const$ trajectories. As can be seen from eq. (\ref{duesedici}) the $\cosh^{-2}$ term has indeed a maximum along the null trajectories $u=u'$ confirming the expectations. There are however also the ``geometrical" prefactors $f^{-2}$ that can mask the above behaviour. 

\section{Acoustic BH model}
\label{s3}

Acoustic BH metrics are mostly characterized by a single sonic point separating the subsonic from the supersonic region of the flow. Both regions are asymptotically homogeneous. A profile mathematically simple enough to manipulate and sufficiently representative is given by a metric for which the conformal factor reads 
\be \label{treuno} f(r)=\tanh 2\kappa r \ , \ee
where $-\infty<r<+\infty$. It has an horizon at $r=0$ and $\kappa=\frac{|f'(r)|_{r=0}}{2}$ is its surface gravity. The subsonic region is $r>0$ while the supersonic one is at $r<0$. 

 \begin{figure}[h]
\centering \includegraphics[angle=0, height=2in] {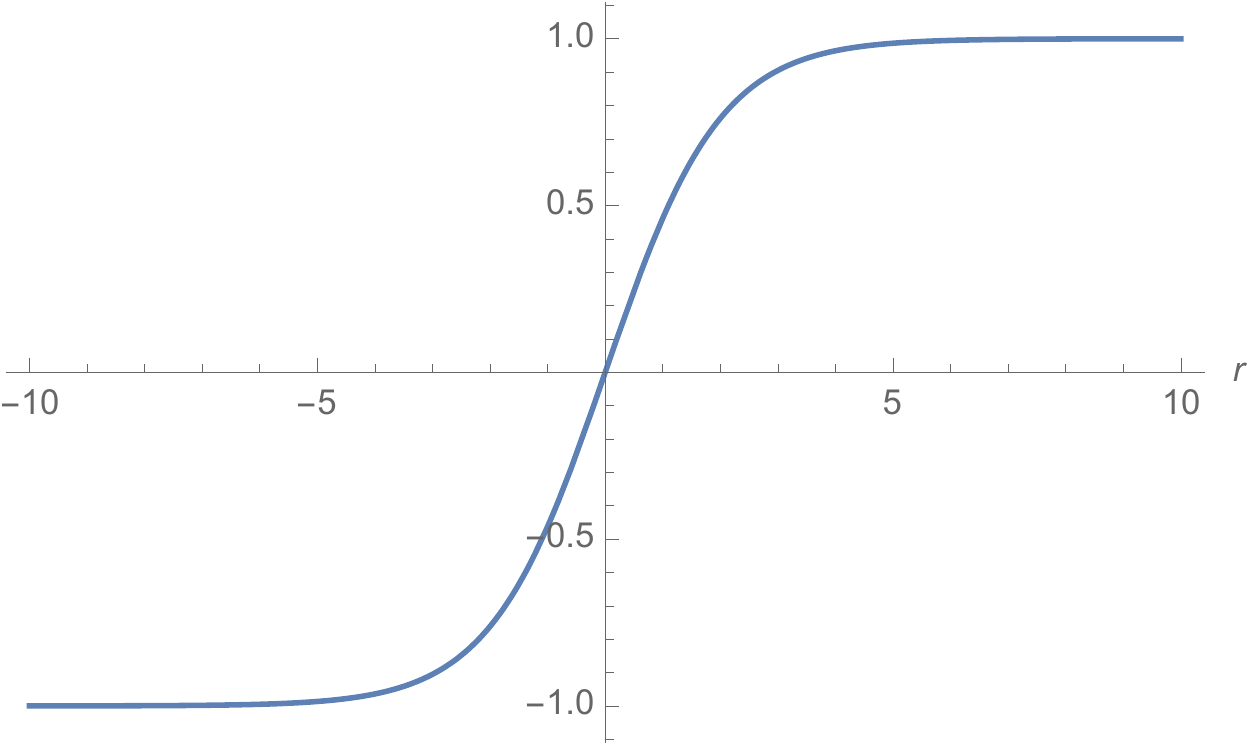}
\caption{The acoustic BH profile (\ref{treuno}) for $\kappa=\frac{1}{4}$.}
\label{figuno}
\end{figure} 

\begin{figure}[h]
\centering \includegraphics[angle=0, height=2in] {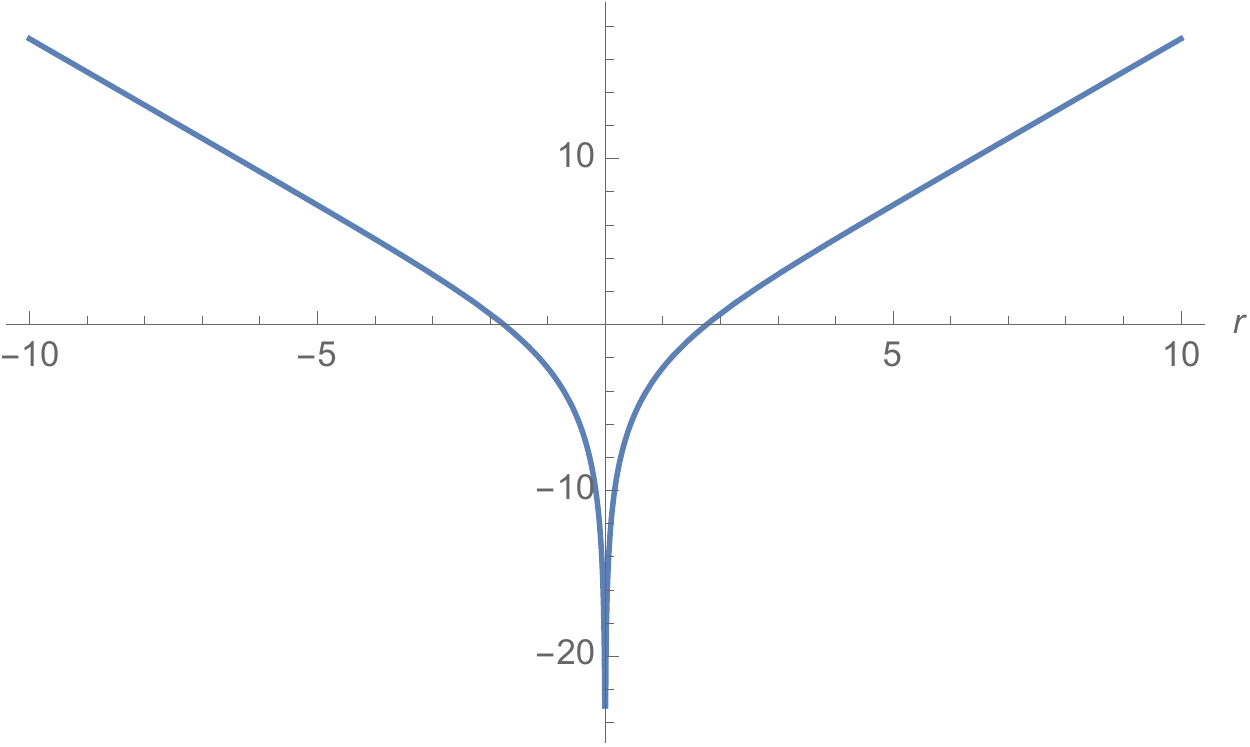}
\caption{Plot of the rhs and the lhs of eq. (\ref{trecinque}) for $\kappa=\frac{1}{4}$.}
\label{figdue}
\end{figure}

In the acoustic language the profile would correspond to a flow with velocity $V(r)$ such that 
\be \label{tredue} f=1-V^2(r) \ . \ee
One see from Fig. (\ref{figuno}) that very rapidly the profile becomes flat indicating an homogeneous flow. From eqs. (\ref{treuno}) and (\ref{duetre}) we have that in this case 
\be \label{tretre} r_*=\frac{1}{2\kappa}\ln|\sinh 2\kappa r| \ , \ee
and from eq. (\ref{duedue})
\be \label{trequattro} 
u=v-\kappa^{-1} \ln|\sinh 2\kappa r|\ . \ee
 The condition for the maximum of the $\cosh^{-2}$ term in $G(x,x')$ at equal $v$ is (see eq. (\ref{duediciassette})) 
 \be \label{trecinque} 
 \kappa^{-1} \ln|\sinh 2\kappa r|=\kappa^{-1} \ln|\sinh 2\kappa r'| \ , \ee
 where $r>0$ and $r'<0$, and is plotted in Fig. (\ref{figdue}).
 
\begin{figure}[h]
\centering \includegraphics[angle=0, height=2in] {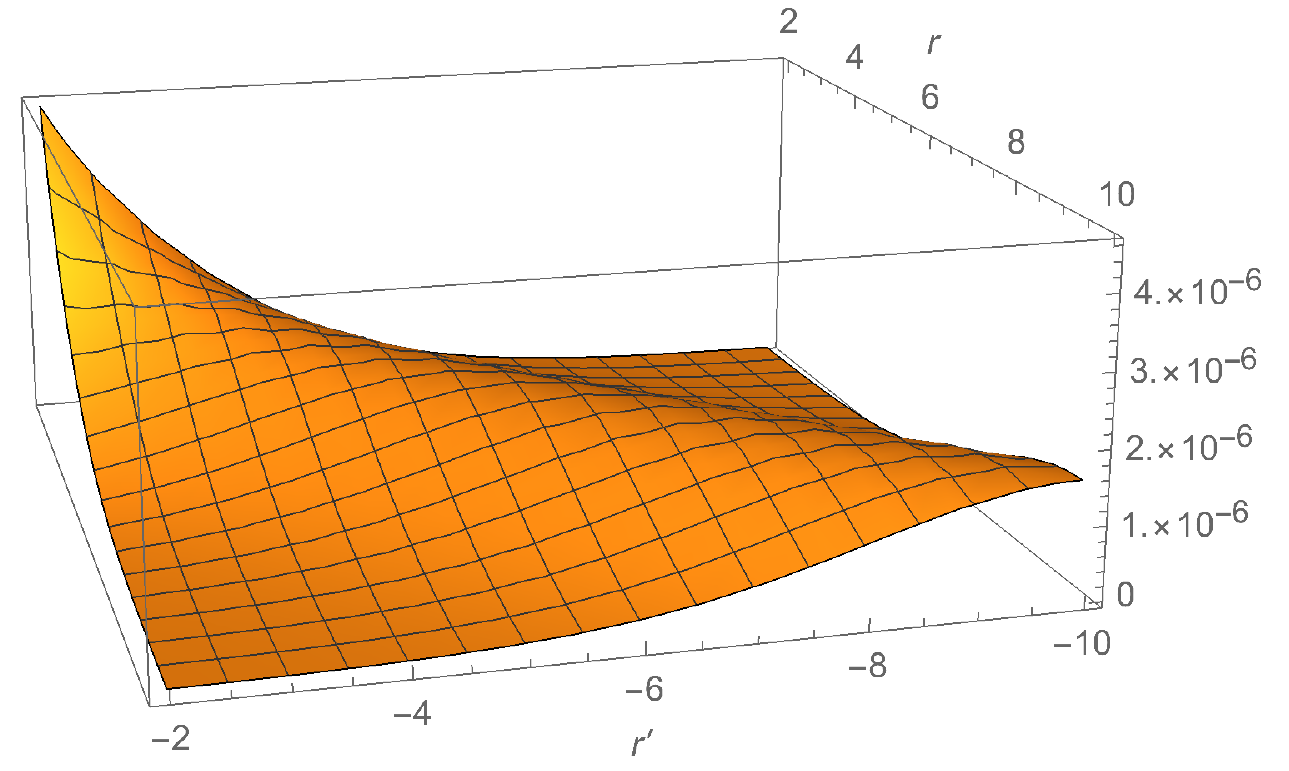}
\caption{Plot of the correlator $G(r,r')$, eq. (\ref{duesedici}) for the acoustic BH model, for $\hbar=1$ and $\kappa=\frac{1}{4}$.}
\label{figtre}
\end{figure}

\begin{figure}[h]
\centering \includegraphics[angle=0, height=2in] {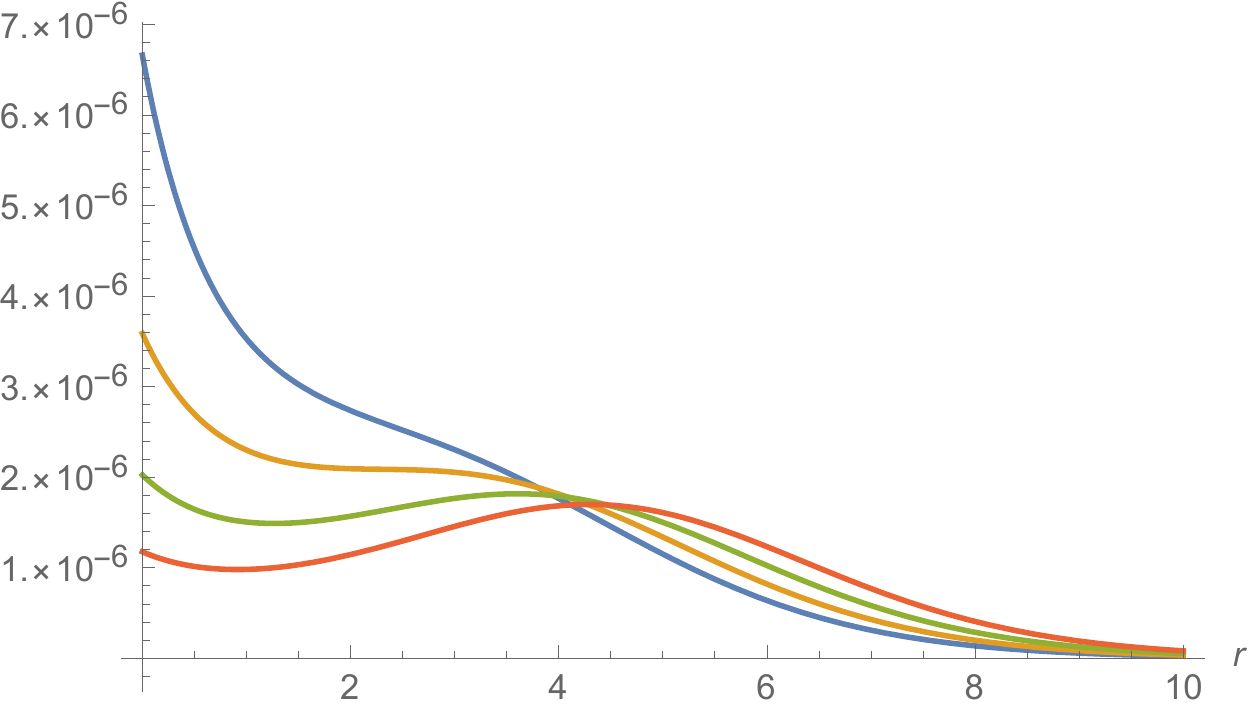}
\caption{Plot of $G(r,r')$ for the acoustic BH model for $\hbar =1$ and $\kappa=\frac{1}{4}$, as a function of $r$ and fixed $r'=-3$ (blue line), $r'=-3.5$ (orange), $r'=-4$ (green) and $r'=-4.5$ (red).}
\label{figquattro}
\end{figure}

In Fig. (\ref{figtre}) a plot of $G(r,r')$ is shown. 
One clearly sees the presence of the expected peak signaling the correlations between the Hawking particles and their partners. The location of the peak is indeed along eq. (\ref{trecinque}) for sufficiently large $r$ (and $r'$) where the prefactors $f^{-2}$ approach rapidly one. For points near the horizon the situation is different. To see what happens there, in Fig. (\ref{figquattro}) we plot $G(r,r')$ as a function of $r$ for vaious fixed values of $r'$.

The peak appears at $u=u'|_{v=v'}$ only for points $r,|r'|\gtrsim \frac{1}{\kappa}$. For points located closer to the horizon no peak appears \cite{schutzholdunruh}, the maximum disappears and it merges in the light-cone singularity at coincidence points (i.e. $r=|r'|=0$). This behaviour corroborates the idea that the Hawking particle and its corresponding partner emerge on shell out of a region of nonvanishing extension across the horizon called ``quantum atmosphere" \cite{qatm1,qatm3, qatm4}.
 In this case it has an extension of order $\frac{1}{\kappa}$. Inside this quantum atmosphere vacuum polarization and Hawking radiation are comparable and one cannot disentangle the two. 
 
 \section{Schwarzschild BH}
\label{s3}

\begin{figure}[h]
\centering \includegraphics[angle=0, height=2in] {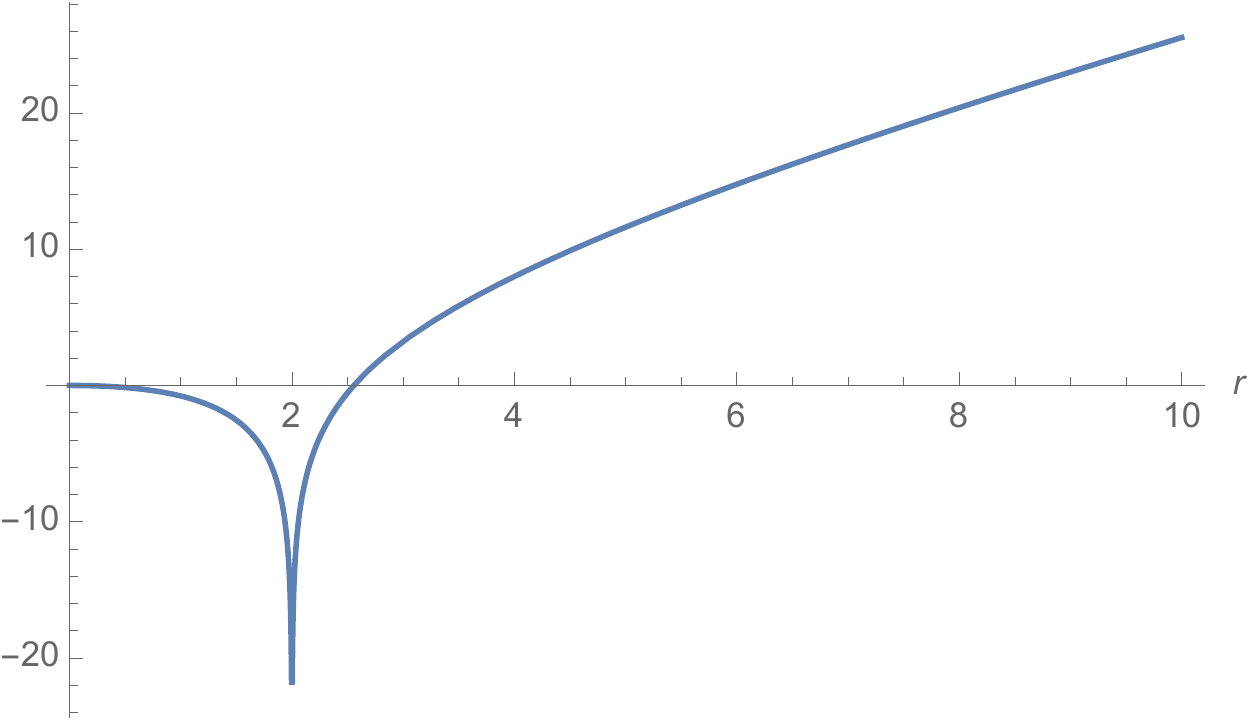}
\caption{Plot of both sides of eq. (\ref{quattroquattro}) for $m=1$.}
\label{figcinque}
\end{figure}

The Schwarzschild BH is characterized by 
\be \label{quattrouno}
f(r)=1-\frac{2m}{r} \ , \ee
where $m$ is the mass of the BH and $r>0$. The horizon is at $r_h=2m$, its surface gravity is $\kappa=\frac{1}{4m}$, while $r=0$ is the physical singularity. 
In this case
\be \label{quattrodue}  r_*=r+2m\ln|\frac{r}{2m}-1|\ , \ee
 and
 \be \label{quattrotre} u=v-2r-4m\ln|\frac{r}{2m}-1|\ . \ee
 The condition $u=u'|_{v=v'}$ is shown in Fig. (\ref{figcinque}) and it reads
 \be \label{quattroquattro} 2r+4m\ln|\frac{r}{2m}-1| = 2r'+4m\ln|\frac{r'}{2m}-1|  \ , 
 \ee

\begin{figure}[h]
\centering \includegraphics[angle=0, height=2in] {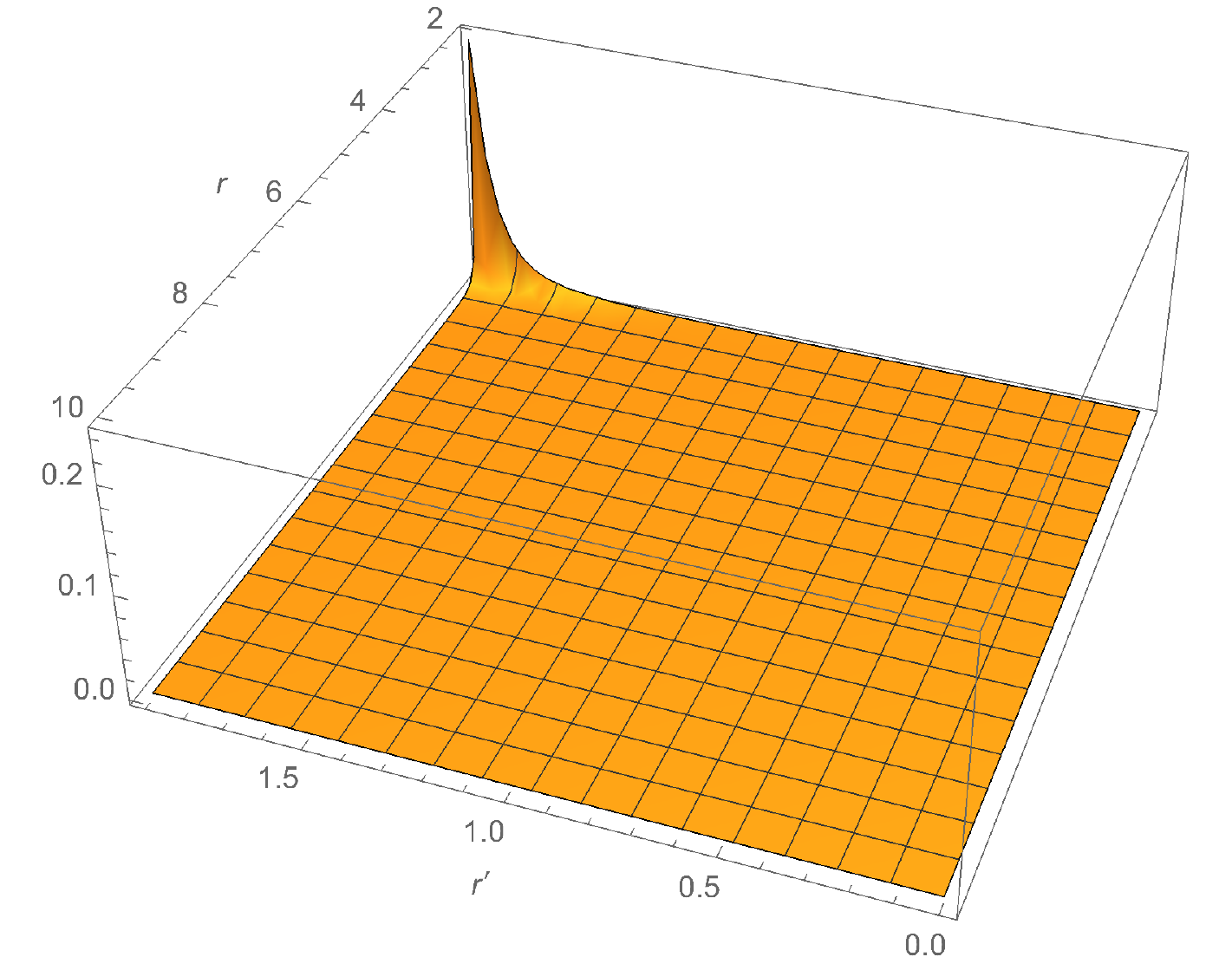}
\caption{
3D plot of $G(r,r')$ for Schwarzschild ($\hbar=1,\ m=1$).
}
\label{figsei}
\end{figure}

The 3D plot of the correlator $G(r,r')$ is given in Fig. (\ref{figsei}). 

\begin{figure}[h]
\centering \includegraphics[angle=0, height=2in] {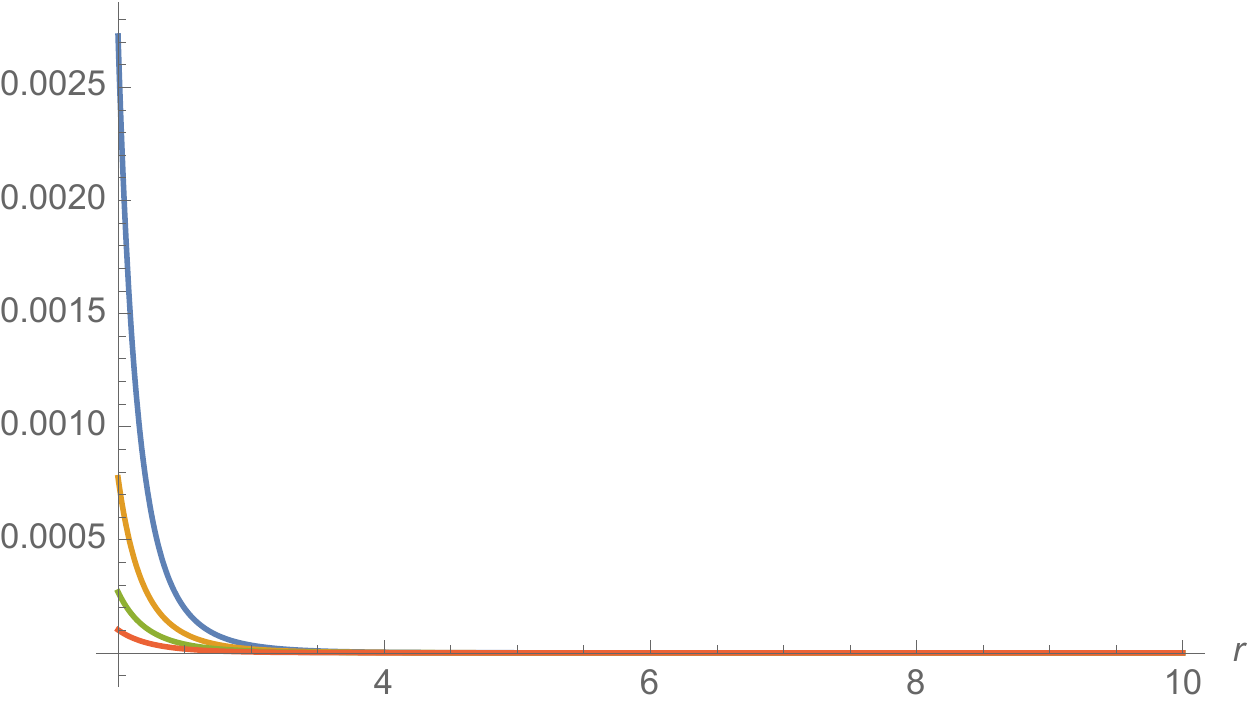}
\caption{
Plots of  $G(r,r')$ for Schwarzschild  ($\hbar=1, m=1$) as a function of $r$ for fixed values of $r'=1.4m$ (blue line), $1.2m$ (orange), $m$ (green), $0.8m$ (red).
}
\label{figsette}
\end{figure}

\begin{figure}[h]
\centering \includegraphics[angle=0, height=2in] {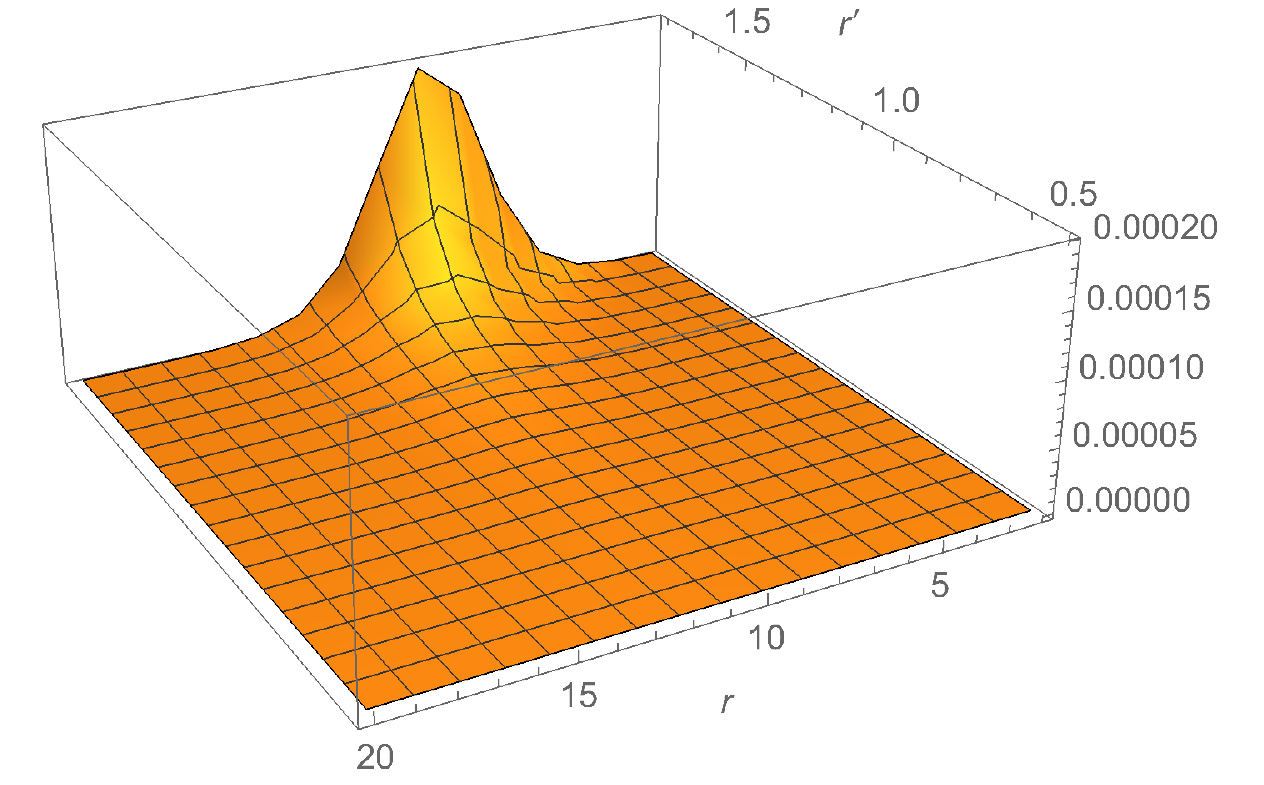}
\caption{3D Plot of  $G(v,r;v',r')$ for Schwarzschild ($\hbar=1, m=1$) and $v-v'=30m$.}
\label{figotto}
\end{figure}

\begin{figure}[h]
\centering \includegraphics[angle=0, height=2in] {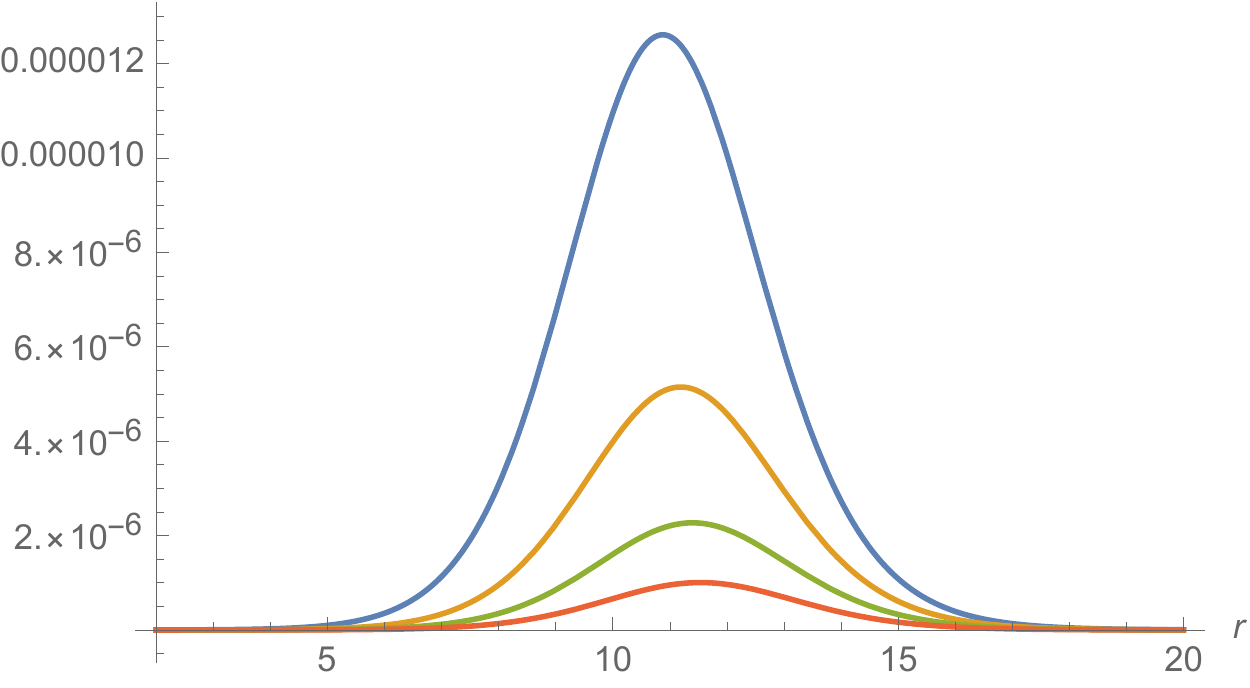}
\caption{Plots of  $G(v,r;v',r')$ for Schwarzschild ($\hbar=1, m=1$), $v-v'=30m$, as a function of $r$ for fixed values of $r'=1.4m$ (blue line), $1.2m$ (orange), $m$ (green), $0.8m$ (red).}
\label{fignove}
\end{figure}

One does not see any structure \cite{Balbinot:2021bnp}. The expected peak does not show up. This can also be seen from Fig. (\ref{figsette}) where $G(r,r')$ is plotted as a function of $r$ for various values of $r'$. 
The reason of this negative result is not that correlations between the Hawking particles and their partners do not exist in this case, simply they do not show up in the equal time correlators. Looking at Fig. (\ref{figcinque}), one sees that the solution of eq. (\ref{quattroquattro}) exists only for a tiny interval of values of $r$ outside the horizon where the vacuum contribution is not negligible. On the other hand, when the Hawking particle emerges on shell out of the vacuum fluctuation, the corresponding particle has already been swallowed by the singularity and correlations do not show up. To reveal them one has to intercept the partner before it gets swallowed by the singularity. To do that the correlator has to be computed not at equal times $v=v'$ but at $v'\ll v$. In Fig. (\ref{figotto}) we plot $G(v,r;v',r')$ for $v-v'=30m$ and the same correlator as a function of $r$ for several fixed values of $r'$ in Fig. (\ref{fignove}). The peak indeed appears along $u=u'$. 

 \section{The CGHS BH}
\label{s4}

\begin{figure}[h]
\centering \includegraphics[angle=0, height=2in] {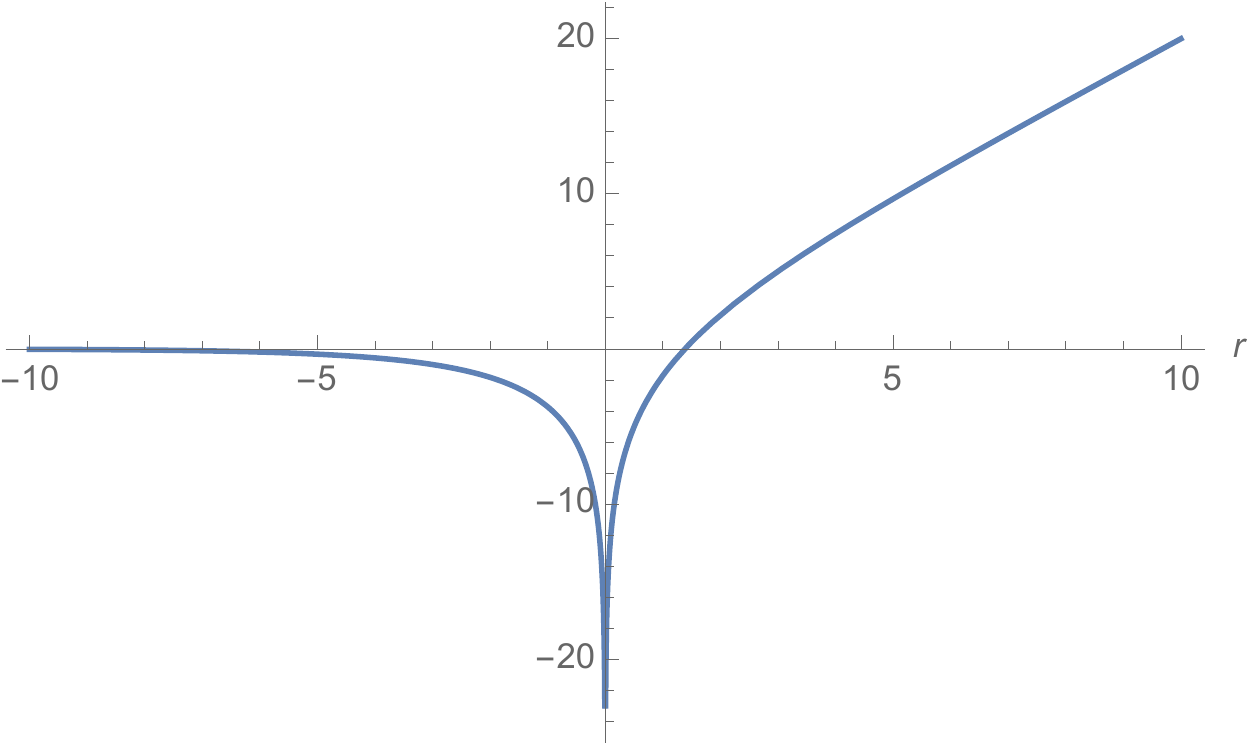}
\caption{Plot of both sides of (\ref{cinquecinque}) for $m=\lambda=\frac{1}{4}$.}
\label{figdieci}
\end{figure}

We have seen that in a BH in order to see the correlations one has to catch the partner before it reaches the singularity. Here we consider a BH metric where the singularity is pushed at $r=-\infty$. 
The Callan-Giddings-Harvey-Strominger BH we shall discuss appears as a solution of a 2D dilaton gravity theory (see \cite{Callan:1992rs} for details) and its metric reads
\be \label{cinqueuno}
ds^2=-(1-\frac{m}{\lambda}e^{-2\lambda r})dv^2+2dvdr \ , \ee
where $m$ is the mass of the BH and $\lambda$ is a parameter interpreted as a cosmological constant. 
Here $-\infty < r < +\infty$. The horizon is located at 
\be \label{cinquedue} r_h=\frac{1}{2\lambda}\ln \frac{m}{\lambda} \ee
and the corresponding surface gravity is $\kappa=\lambda$. The metric has a physical singularity at
$r=-\infty$ where the curvature diverges. For this metric
\be \label{cinquetre} u=v-2r_*\ , \ee
 where now
 \be \label{cinquequattro} r_*=r+\frac{1}{2\lambda}\ln|1-\frac{m}{\lambda}e^{-2\lambda r}| \ee
  and the condition $u=u'|_{v=v'}$ reads
  \be \label{cinquecinque} 2r+\frac{1}{\lambda}\ln|1-\frac{m}{\lambda}e^{-2\lambda r}| = 2r'+\frac{1}{\lambda}\ln|1-\frac{m}{\lambda}e^{-2\lambda r'}|\ .  \ee
 Here $r>r_h$ and $r'<r_h$. In Fig. (\ref{figdieci}) we plot both sides of eq. (\ref{cinquecinque}) . Note that the value of $r$ corresponding  to $r'\to -\infty$   is
 \be \label{cinquesei} r|_{r'=-\infty}=\frac{1}{2\lambda}\ln 2 + r_h \ . \ee
 So $r'=-\infty$ is correlated to a point still inside the quantum atmosphere. 

\begin{figure}[h]
\centering \includegraphics[angle=0, height=2in] {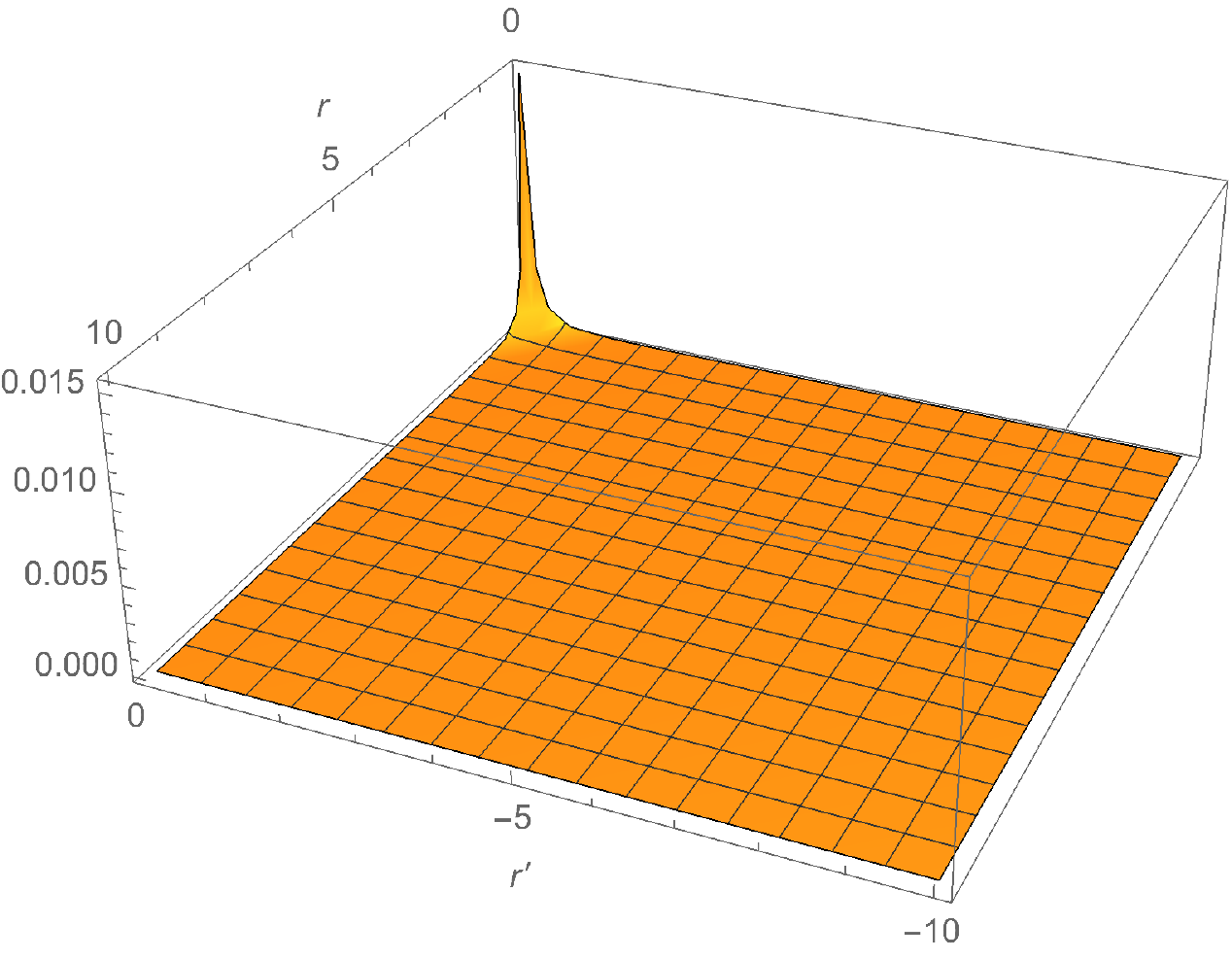}
\caption{Plot of $G(r,r')$ for the CGHS BH, $\hbar=1$ and $m=\lambda=\frac{1}{4}$.}
\label{figundici}
\end{figure}

As shown in Fig. (\ref{figundici}), when the correlator $G(r,r')$ for this metric is plotted, no correlation shows up. The explanation of this negative result relies on the fact that, although the singularity is at $r=-\infty$, it is not ``infinitely" far away. The proper distance
\be \label{cinquesei} s=\int_{r_0}^{-\infty}\frac{dr}{(1-\frac{m}{\lambda}e^{-2\lambda r})^{\frac{1}{2}}} \ee
is finite. 
 
The CGHS metric shown has therefore the same behaviour we found in the Schwarzschild metric concerning the correlation across the horizon. The partner is swallowed by the singularity before the Hawking particle emerges out of the quantum atmosphere.  
 
\section{Simpson-Visser metric}
\label{s5}

\begin{figure}[h]
\centering \includegraphics[angle=0, height=3.5in] {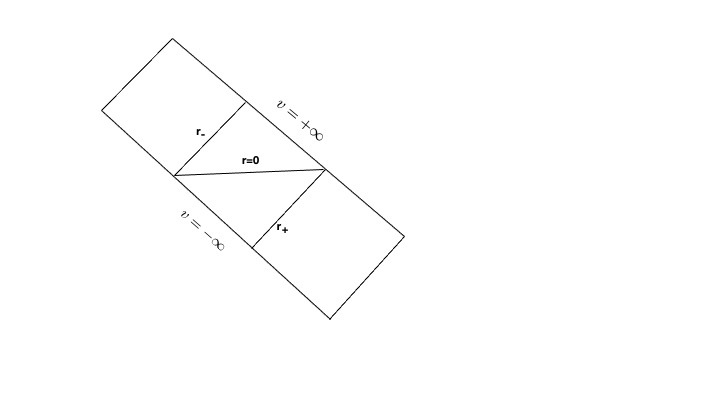}
\caption{
Penrose diagram of the part of the Simpson-Visser metric covered by the $v,r$ coordinates.
}
\label{figdodici}
\end{figure}

Our last example concerns a BH metric where the singularity has been removed, one has a ``regular BH". Among many proposals in the literature for this kind of BHs, we confine our attention to a very simple metric, the so called Simpson-Visser metric \cite{Simpson:2018tsi}, for which
\be \label{seiuno} ds^2=-(1-\frac{2m}{\sqrt{r^2+a^2}})dv^2+2dvdr \ , \ee
where $a$ is a parameter, regularizing the singularity, which we choose such that $a<2m$. For $a=0$ we have the Schwarzschild metric with singularity at $r=0$. For $a\neq 0$ the spacetime surface $r=0$ is regular and represents a bounce which separates one asymptotically flat Universe (where $r>0$) from an identical copy with $r<0$. 
\be \label{seidue} r_{\pm}=\pm\sqrt{(2m)^2-a^2} \ee
corresponds to a pair of horizons ($|a|<2m$). The part of the Penrose diagram of the Simpson-Visser metric covered by the $(v,r)$ coordinates is given in Fig. (\ref{figdodici}). 

For this metric 
\bea \label{seitre}
 r_* &=& \int \frac{dr}{1-\frac{2m}{\sqrt{r^2+a^2}}}=r+2M \ln \left( \frac{r}{a}+\sqrt{ \left(\frac{r}{a}\right)^2+1} \right) 
\nonumber \\
&+& \frac{4M^2}{\sqrt{4M^2-a^2}}\ln\frac{ | \sqrt{4M^2-a^2}\tanh \left( \frac{\sinh^{-1} (\frac{r}{a})}{2} \right)     -2M+a |}{ |  \sqrt{4M^2-a^2}\tanh \left( \frac{\sinh^{-1} (\frac{r}{a})}{2} \right)+2M-a |}\ 
 \eea
 and 
 \bea \label{seiquattro}
 u=v-2r_*&=&v - 2r-4M \ln \left( \frac{r}{a}+\sqrt{ \left(\frac{r}{a}\right)^2+1} \right) 
\nonumber \\
&-& \frac{8M^2}{\sqrt{4M^2-a^2}}\ln\frac{ | \sqrt{4M^2-a^2}\tanh \left( \frac{\sinh^{-1} (\frac{r}{a})}{2} \right)     -2M+a |}{ |  \sqrt{4M^2-a^2}\tanh \left( \frac{\sinh^{-1} (\frac{r}{a})}{2} \right)+2M-a |}\ .
 \eea
 
\begin{figure}[h]
\centering \includegraphics[angle=0, height=3.5in] {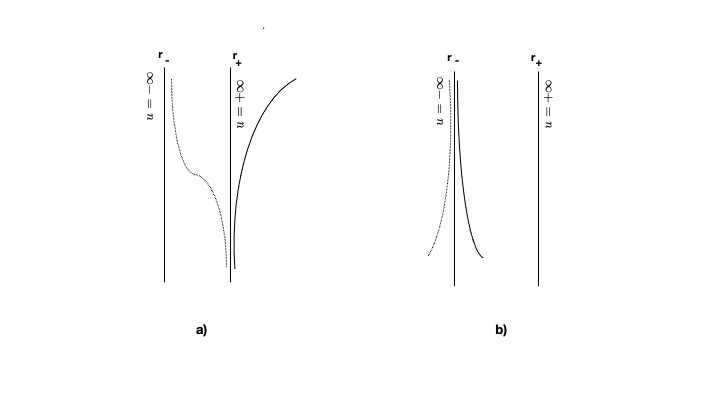}
\caption{
a) Trajectories of a Hawking particle (solid line) and partner (dashed line) pair created near $r_+$; b) pair creation near $r_-$.
}
\label{figtredici}
\end{figure}

 Fig. (\ref{figtredici}a) represents the trajectories of a Hawking particle (solid line) - partner (dashed line) pair created near $r_+$, while Fig. (\ref{figtredici}b) a pair created near $r_-$.  All particles propagate along $u=const$.
One can see that in the first case ((\ref{figtredici}a)), while the Hawking particle goes away to infinity, the partner piles up along $r_-$. In the second case, both the Hawking particle and the partner pile up along $r_-$. This situation remembers what happens in the interior of the Reissner-Nordstr\"om BH. As in that case one has to introduce two Kruskal coordinates, one regular on $r_+$
\be \label{seicinquea} U_{(+)}=\mp\frac{1}{\kappa}e^{-\kappa u}\ , \ee
where $-$ refers to $r>r_+$ and $+$ to $r<r_+$; the other regular on $r_-$ 
\be \label{seicinqueb} U_{(-)}=\pm \frac{1}{\kappa}e^{\kappa u}\ , \ee
where $+$ refers to $r>r_-$ and $-$ to $r<r_-$. These $\kappa$ is the absolute value of the surface gravity, which is the same for both horizons
\be \label{seisei}
\kappa\equiv \frac{1}{2} |\frac{df}{dr}|_{r_{\pm}}=\frac{\sqrt{4M^2-a^2}}{8M^2} \ .\ee
Note that the coordinate $U_{(+)}$  is regular on $r_+$, where $U_{(+)}=0$, but is singular on $r_-$, where $U_{(+)}=+\infty$. 

We can define a Unruh state $|U_{(+)}\rangle$ expanding the quantum fields in modes like (\ref{duedodici}) where $U=U_{(+)}$. Similarly, $|U\_{(-)}\rangle$  is defined by the expansion (\ref{duedodici}) where now $U=U_{(-)}$. The singularities of the coordinates (\ref{seicinquea}), (\ref{seicinqueb}) induce singularities in the corresponding modes, for example $e^{-i\omega U_{(+)}}$ is singular on $r_-$. However, being the surface gravity the same in absolute value one can show that the quantum stress tensor is the same in both states $|U_{(\pm)}\rangle$ and is regular on both horizons (see appendix \ref{appendixA}). Also the correlator, being an even function of the surface gravity (see eq. (\ref{duesedici})), is the same in $|U_{(\pm)}\rangle$. 

\begin{figure}[h]
\centering \includegraphics[angle=0, height=2.in] {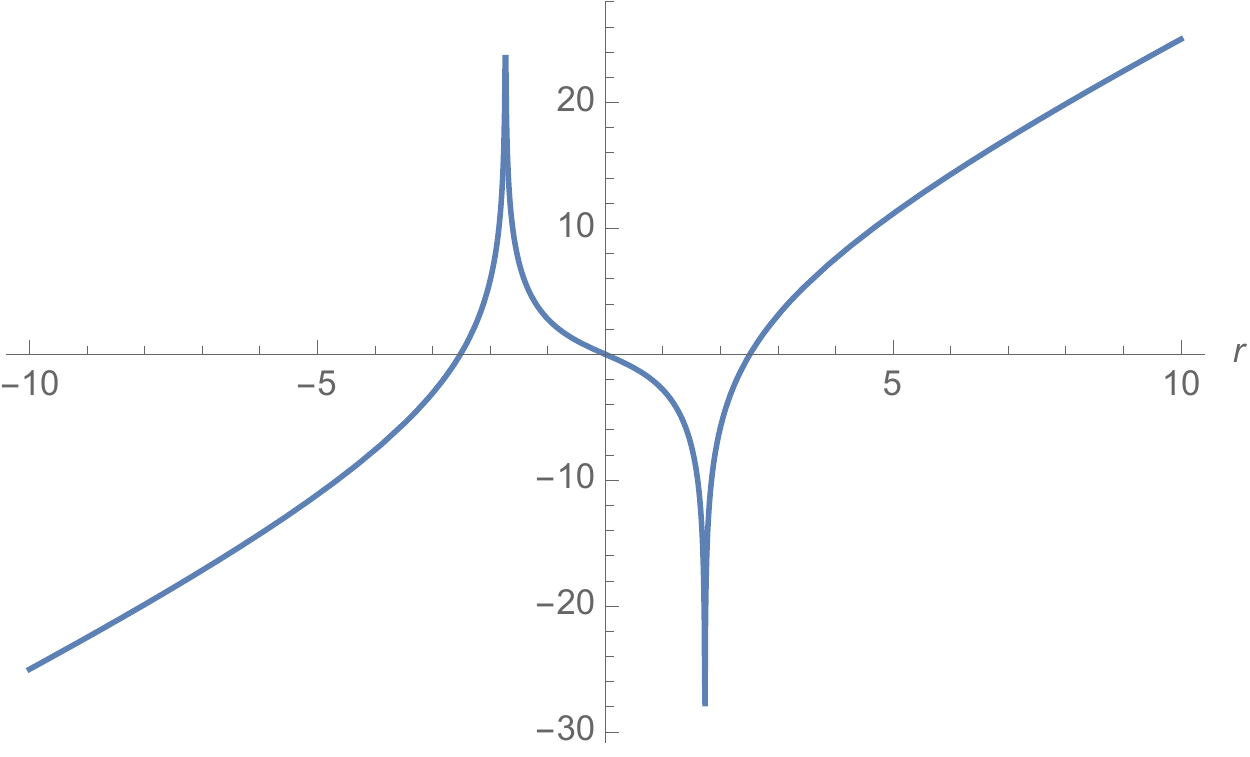}
\caption{
Plot of both sides of eq. (\ref{seisette}), where $a=1,M=1$, $r_\pm =\pm\sqrt{3}\sim \pm 1.732,\ \kappa=\frac{\sqrt{3}}{8}\sim 0.217$.
}
\label{figquattordici}
\end{figure}

The extremal of the $\cosh^{2}$ term is now given by $u=u'|_{v=v'}$ which reads
 \bea \label{seisette}
 && 2r+4M \ln \left( \frac{r}{a}+\sqrt{ \left(\frac{r}{a}\right)^2+1} \right) 
+ \frac{8M^2}{\sqrt{4M^2-a^2}}\ln\frac{ | \sqrt{4M^2-a^2}\tanh \left( \frac{\sinh^{-1} (\frac{r}{a})}{2} \right)     -2M+a |}{ |  \sqrt{4M^2-a^2}\tanh \left( \frac{\sinh^{-1} (\frac{r}{a})}{2} \right)+2M-a |} = \nonumber \\
&& 2r'+4M \ln \left( \frac{r'}{a}+\sqrt{ \left(\frac{r'}{a}\right)^2+1} \right) 
+ \frac{8M^2}{\sqrt{4M^2-a^2}}\ln\frac{ | \sqrt{4M^2-a^2}\tanh \left( \frac{\sinh^{-1} (\frac{r'}{a})}{2} \right)     -2M+a |}{ |  \sqrt{4M^2-a^2}\tanh \left( \frac{\sinh^{-1} (\frac{r'}{a})}{2} \right)+2M-a |}, \ \ \ \ \ \  \eea
plotted in Fig. (\ref{figquattordici}),
where $r>r_+$ and $r'<r_+$ for the case of Fig. (\ref{figtredici}a) while $r<r_-$ and $r'>r_-$ for Fig. (\ref{figtredici}b).\footnote{From Figs. (\ref{figtredici}a) and (\ref{figquattordici})  one sees that the condition (\ref{seisette}) is always satisfied, but the points with $r-r_+\gtrsim \frac{1}{\kappa}$ are correlated with corresponding partners that are piling up close to $r_-$.} 

\begin{figure}[h]
\centering \includegraphics[angle=0, height=2.in] {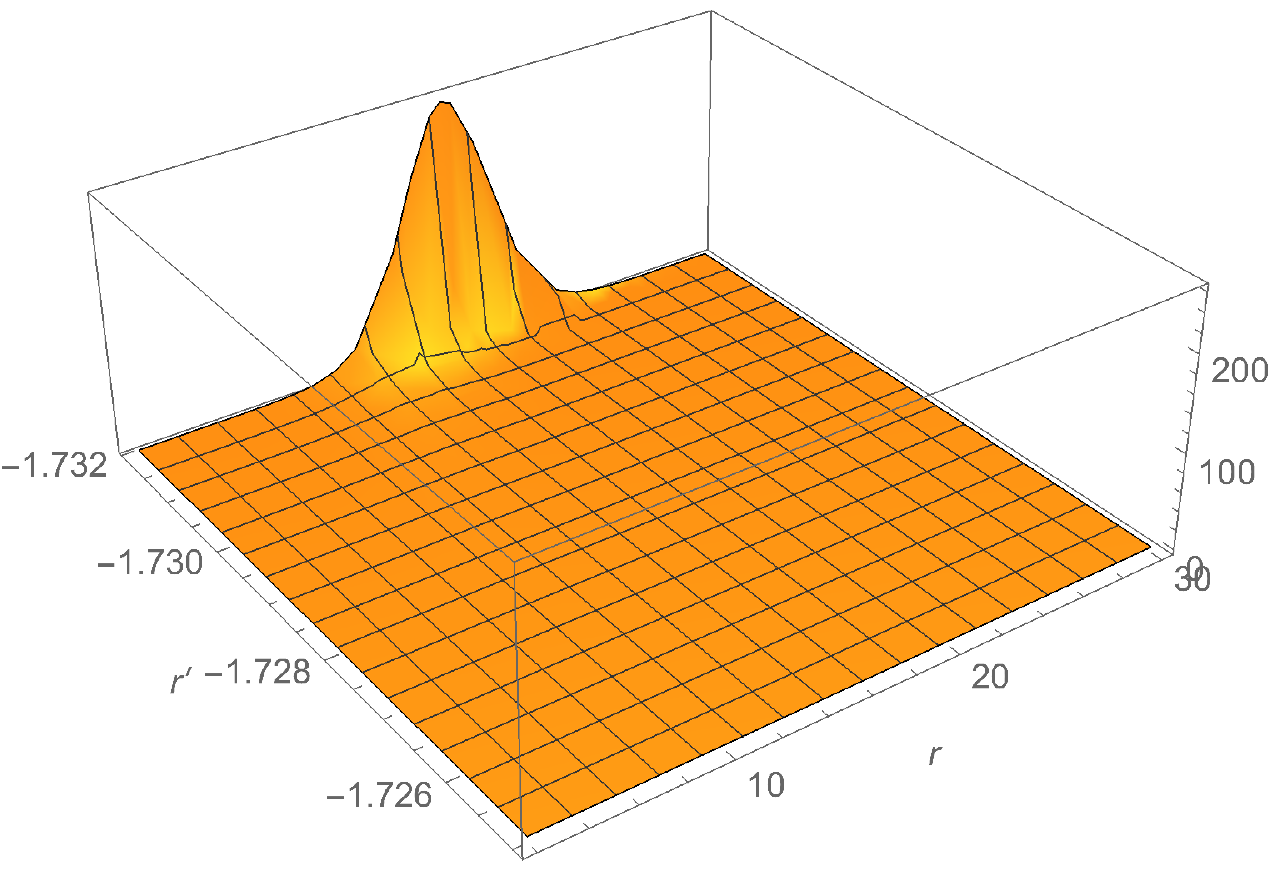}
\caption{
Plot of  $G(r,r')$ for the Simpson-Visser BH, $\hbar=1$ and $a=1,M=1$, $r_\pm =\pm\sqrt{3}\sim \pm 1.732$.
}
\label{figquindici}
\end{figure}

\begin{figure}[h]
\centering \includegraphics[angle=0, height=2.in] {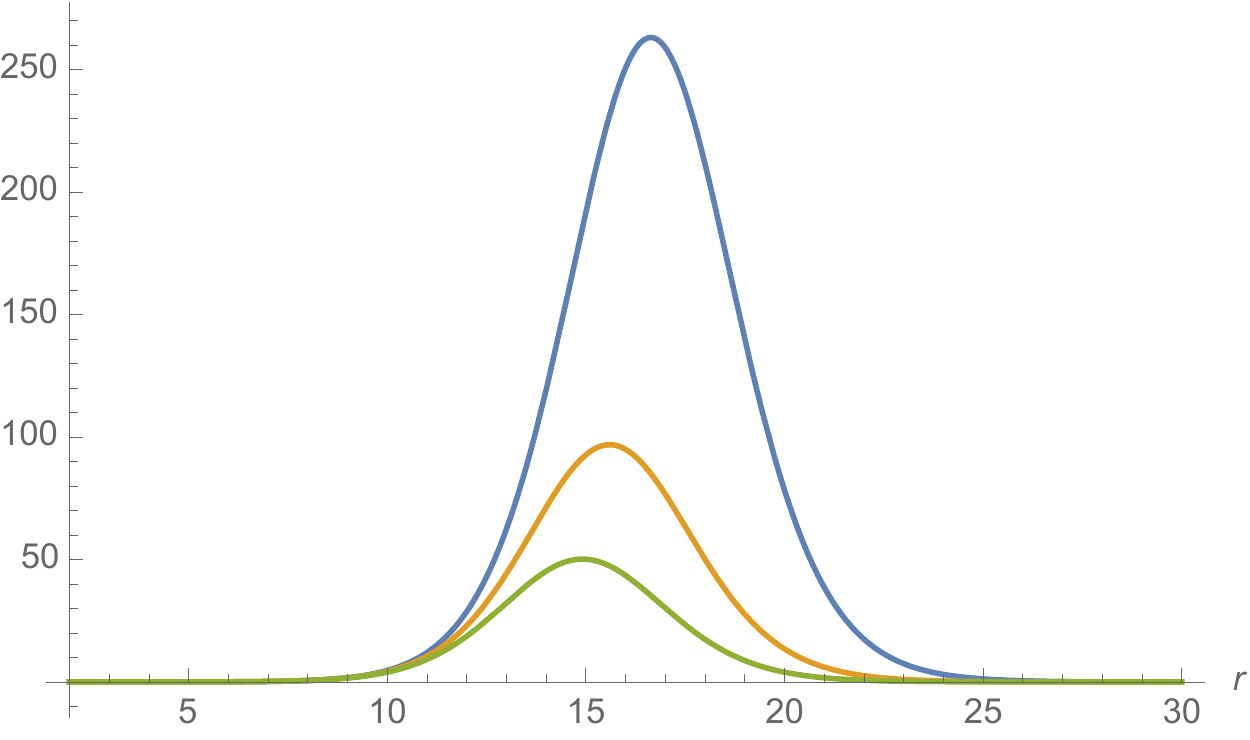}
\caption{
Plot of  $G(r,r')$ for the Simpson-Visser BH, $\hbar=1$ and $a=1,M=1$, $r_\pm =\pm\sqrt{3}\sim \pm 1.732$, as a function of $r$ at fixed values of $r'=-1.7319$ (blue curve), $-1.7318$ (orange), $-1.7317$ (green).
}
\label{figsedici}
\end{figure}

The corresponding correlator\footnote{For a preliminary study see \cite{Fontana:2023zqz}.}
is graphically represented in Fig. (\ref{figquindici}).
One can further appreciate this by examining the correlator at fixed inner point $r'$, as shown in Fig. 
(\ref{figsedici}). The $r,r'$ values for the peak are in good agreement with the expected one, $u=u'$, given, at equal time, by eq. (\ref{seisette}). 

\begin{figure}[h]
\centering \includegraphics[angle=0, height=2.in] {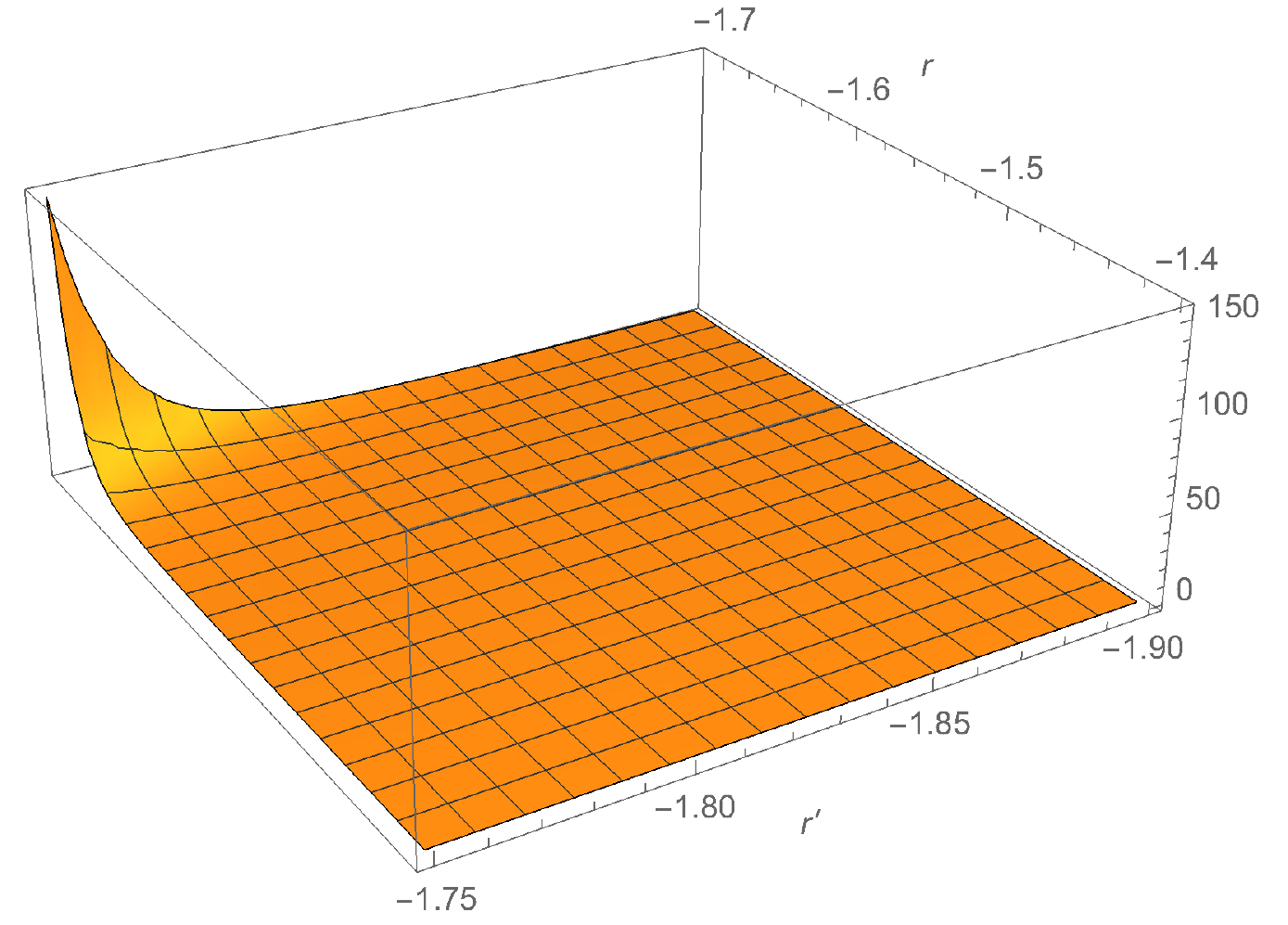}
\caption{
Plot of  $G(r,r')$ for the Simpson-Visser BH, $\hbar=1$ and $a=1,M=1$, $r_\pm =\pm\sqrt{3}\sim \pm 1.732$, for the case of Fig. (\ref{figtredici}b).
}
\label{figdiciassette}
\end{figure}

On the contrary, the correlator corresponding to Fig. (\ref{figtredici}b) does not show any sign of correlations, see Fig. (\ref{figdiciassette}). This is understandable since both the Hawking particle and the partner now pile up along $r_-$ remaining well inside the quantum atmosphere where the correlator is dominated by the coincidence limit singularity.

\section{Conclusion}

Hawking radiation is a genuine pair (particle-partner) production process that is expected to be a general feature of gravitational BHs and of analogue ones realized in condensed matter systems. It is indeed in these latter (and only in them) that the existence of this radiation has been (indirectly) observed. The observation consisted in detecting the correlations across the sonic horizon between the Hawking particles and their entangled partners. 

In this paper we have analyzed various examples of BH spacetime metrics to see if and where these correlations show up. Indeed they can be hidden by the geometrical structure of the underlying spacetime, like the presence of singularities or inner horizons as shown explicitly here. We have seen that singularities (without inner horizons) swallow the partners before the corresponding Hawking particles emerge on shell out of the quantum atmosphere obscuring the existing correlations. On the other hand inner horizons produce a piling up of the partners along them enhancing and strongly localizing the correlations.

\begin{appendix}

\section{}
\label{appendixA}

Consider a 2D space-time metric
\be \label{auno}
ds^2=C(r)dudv \ , \ee
where $u$ and $v$ are null coordinates, usually expressed in terms of a Schwarzschild-like time $t$ as
\bea
u&=& t-r_*\ , \label{aduea} \\
v&=&t+r_* \ , \label{adueb} \eea
and
\be \label{atre} r_*=\int \frac{dr}{C}\ . \ee

$C(r)$ is supposed to vanish at $r\to r_+$ (outer horizon) and at $r=r_-$ (inner horizon). The expectation values of the stress energy tensor operator for a quantum massless scalar field in the Unruh state, defined by expanding the quantum field on the basis 
$\{ \frac{e^{-i\omega_K U_{(+)}}}{\sqrt{4\pi\omega_K}},\ \frac{e^{-i\omega v}}{\sqrt{4\pi\omega}} \}$, where 
\be \label{aquattro} U_{(+)}=\mp \frac{1}{\kappa_+}e^{-\kappa_+ U_{(+)}} \ee
and $\kappa_+$ is the surface gravity of the horizon $r_+$, are \cite{Birrell:1982ix}
\bea \langle U_{(+)}|\hat T_{vv}|U_{(+)}\rangle &=& -\frac{1}{192\pi}\left[ C'(r)^2-2C(r)C''(r)\right]\ , \label{acinquea} \\
\langle U_{(+)}|\hat T_{uv}|U_{(+)}\rangle &=& \frac{1}{96\pi}C(r)C''(r) \ , \label{acinqueb} \\
\langle U_{(+)}|\hat T_{uu}|U_{(+)}\rangle &=& \langle U_{(+)}|T_{vv}|U_{(+)}\rangle -\frac{1}{24\pi}\{ 
U_{(+)},u\} \ , \label{acinquec} \eea
where a prime $'$ indicates derivative with respect to $r$, and $\{ 
U_{(+)},u\}$ is the Schwarzian derivative between $U_{(+)}$ and $u$ which reads
\be \label{asei} -\frac{1}{24\pi}\{ 
U_{(+)},u\}=\frac{\kappa_+^2}{48\pi}\ . \ee
$\langle U_{(+)}| \hat T_{ab} | U_{(+)}\rangle $ is regular on the outer horizon $r=r_+$ ($U_{(+)}=0$) if it is finite in a regular frame there
\bea \lim_{r\to r_+} \frac{\langle U_{(+)}| \hat T_{uu} | U_{(+)}\rangle}{C^2} <\infty \ , \label{asettea} \\
 \lim_{r\to r_+} \langle U_{(+)}| \hat T_{vv} | U_{(+)}\rangle  <\infty \ , \label{asetteb} \\
 \lim_{r\to r_+} \frac{\langle U_{(+)}| \hat T_{uv} | U_{(+)}\rangle}{C} <\infty \ . \label{asettec} \eea
 This implies that the stress tensor is regular in a free falling frame across the horizon. 
 
 One immediately sees that if $C(r)$ is finite on the horizon with its derivatives up to the second, conditions (\ref{asetteb}) and (\ref{asettec}) are satisfied (see eqs. (\ref{acinquea}) and (\ref{acinqueb})). 
 Concerning the $(u,u)$ component (see eq. (\ref{acinquec})) note that to leading order 
 $\langle U_{(+)}| \hat T_{uu} | U_{(+)}\rangle =-\frac{1}{48\pi}\kappa_+^2 + \frac{1}{48\pi}\kappa_+^2 =0$, i.e. vacuum polarization and Hawking radiation cancel out on the horizon. Furthermore, taking the derivative of eq. (\ref{acinquec}) and evaluating it on the horizon
 \be \frac{d}{dr} \langle U_{(+)}| \hat T_{uu} | U_{(+)}\rangle =-\frac{1}{192\pi}\{ 2C'C''-2C'C''-2CC'' \}|_{r_+}=0 \ . \label{aotto} \ee
 So $ \langle U_{(+)}| \hat T_{uu} | U_{(+)}\rangle$ vanishes on the horizon as $C^2$ and condition (\ref{asettea}) is also satisfied. 
 
 If there is another horizon (the inner one at $r_-$) regularity there requires conditions similar to (\ref{asettea}, \ref{asetteb}, \ref{asettec}) to be satisfied; one simply has to consider the limit $r\to r_-$. No problem with (\ref{asetteb}) and (\ref{asettec}) arises while concerning (\ref{asettea}) we have from (\ref{acinquec}) 
 \be \label{anove} 
  \langle U_{(+)}| \hat T_{uu} | U_{(+)}\rangle|_{r_-}=-\frac{1}{48\pi} ( \kappa_-^2 - \kappa_+^2)\ , 
\ee      
which is nonvanishing whenever $\kappa_+^2\neq \kappa_-^2$ as for a non extremal Reissner-Nordstr\"om BH for example. This makes condition (\ref{asettea}) not satisfied and hence $\langle U_{(+)}| \hat T_{uu} | U_{(+)}\rangle$ is singular on $r_-$. 
One can define an Unruh vacuum with respect to the Kruskal coordinate on $r_-$, i.e.
\be \label{adieci}
U_{(-)}=\mp \frac{1}{\kappa_-}e^{\kappa_- u} \ee
and similarly one finds regularity on $r_-$, but on $r_+$ 
\be \label{aundici} 
  \langle U_{(-)}| \hat T_{uu} | U_{(-)}\rangle|_{r_-}=-\frac{1}{48\pi} ( \kappa_+^2 - \kappa_-^2)\ , 
\ee 
making $\langle U_{(-)}| \hat T_{uu} | U_{(-)}\rangle$ singular on $r_+$. 

In the case of the Simpson-Visser metric we have indeed two horizons, the outer one at $r_+=\sqrt{(2m)^2-a^2}$, but $|\kappa_+|=|\kappa_-|=\kappa$ for this spacetime.
So we have now that both $\langle U_{(+)}| \hat T_{uu} | U_{(+)}\rangle$ and 
$\langle U_{(-)}| \hat T_{uu} | U_{(-)}\rangle$ are indeed regular both
on $r_+$ ($U_{(+)}=0$) and $r_-$ ($U_{(-)}=0$). However a singularity appears on the Cauchy horizon $r=r_-$, $v=+\infty$. Regularity there would require

\bea  \\
 \lim_{r\to r_-} \langle U_{(+)}| \hat T_{uu} | U_{(+)}\rangle  <\infty \ , \label{adodicia} \\
 \lim_{r\to r_-} \frac{\langle U_{(+)}| \hat T_{uv} | U_{(+)}\rangle}{C} <\infty \ , \label{adodicib}\\ 
 \lim_{r\to r_-} \frac{\langle U_{(+)}| \hat T_{vv} | U_{(+)}\rangle}{C^2} <\infty \ . \label{adodicic}   \eea
This last is not satisfied since
\be \label{atredici} 
\lim_{r\to r_-} \langle U_{(+)}| \hat T_{vv} | U_{(+)}\rangle =-\frac{\kappa_-^2}{48\pi}\ . \ee
If on the other hand one constructs the Hartle-Hawking states 
\be \label{aquattordici}
\{ \frac{e^{-i\omega_K U_{(\pm)}}}{\sqrt{4\pi\omega_K}}\ , \frac{e^{-i\omega_K V_{(\pm)}}}{\sqrt{4\pi\omega_K}} \} \ , 
\ee
where
\be \label{aquindici} V_{(\pm)}=\pm \frac{1}{\kappa_{\pm}}e^{\pm\kappa_{\pm} v}\ , \ee
one finds that these states are regular all over the spacetime and describe a BH in thermal equilibrium with thermal radiation at the Hawking temperature
\be \label{asedici} T_H=\frac{\hbar\kappa}{2\pi k_B}\ . \ee

\end{appendix}
 
 \acknowledgments
 
A.F. acknowledges partial financial support by the Spanish Ministerio de Ciencia e Innovaci\'on Grant No. PID2020–116567 GB-C21 funded
by Grant No. MCIN/AEI/10.13039/501100011033, and
the Project No. PROMETEO/2020/079 (Generalitat
Valenciana).

\end{document}